\documentclass[%
 reprint,
 amsmath,amssymb,
aps
pra,twocolumn,
]{revtex4-2}
\usepackage{braket}
\usepackage{graphicx}
\usepackage{bm}
\usepackage{float}
\usepackage{color}
\usepackage{soul}
\usepackage[caption = false]{subfig}
\usepackage{hyperref}
\usepackage{placeins}

\usepackage[normalem]{ulem}

\begin{document}

\title{Unifying S\o rensen-M\o lmer gate and Milburn gate with an optomechanical example}
\author{Yue Ma$^1$}

\author{Manuel C. C. Pace$^1$}

\author{M. S. Kim$^1$}

\affiliation{$^1$QOLS, Blackett Laboratory, Imperial College London, London SW7 2AZ, United Kingdom\\
}

\begin{abstract}

S\o rensen-M\o lmer gate and Milburn gate are two geometric phase gates, generating nonlinear self-interaction of a target mode via its interaction with an auxiliary mechanical mode, in the continuous  and pulsed interaction regime, respectively. In this paper, we aim at unifying the two gates by demonstrating that S\o rensen-M\o lmer gate is the continuous limit of Milburn gate, emphasising the geometrical interpretation in the mechanical phase space. We explicitly consider imperfect gate parameters, focusing on relative errors in time for S\o rensen-M\o lmer gate and in phase angle increment for Milburn gate. We find that, although the purities of the final states increase for the two gates upon reducing the interaction strength together with traversing the mechanical phase space multiple times, the fidelities behave differently. We point out that, the difference exists because the interaction strength depends on the relative error when taking the continuous limit from the pulsed regime, thereby unifying the mathematical framework of the two gates. We demonstrate this unification in the example of an optomechanical system, where mechanical dissipation is also considered. We highlight that, the unified framework facilitates the new method of deriving the dynamics of the continuous interaction regime without solving differential equations.

\end{abstract}


\maketitle

\section{introduction}

Utilising geometric phase~\cite{agarwal1990berry,chaturvedi1987berry} in quantum computation was first proposed~\cite{milburn1999simulating,sorensen1999quantum,sorensen2000entanglement,milburn2000ion} and experimentally realised~\cite{sackett2000experimental,leibfried2003experimental} in the platform of trapped ions, making use of the accumulated phase that equals to the enclosed area of the closed loop traversed in a phase space satisfying the basic quantum mechanical commutator $[\hat{X},\hat{P}]=i$~\cite{luis2001quantum}. Due to its ability to generate effective interaction between non-interacting subsystems~\cite{proctor2016hybrid}, the idea has been widely applied for simulating qubit gates in a variety of physical systems, including in quantum optics~\cite{van2008hybrid} and in superconducting circuits~\cite{spiller2006quantum}. Geometric phase has also been exploited in optomechanical systems, where a bosonic optical field mode interacts with a mechanical oscillator via radiation pressure~\cite{aspelmeyer2014cavity}. The effective nonlinear self-interaction induced by geometric phase facilitates generation of nonclassical field~\cite{bose1997preparation} and oscillator~\cite{khosla2013quantum} states, and even the detection of potential quantum gravitational effects~\cite{pikovski2012probing}.

Among the first proposals of geometric phase gate, S\o rensen-M\o lmer gate~\cite{sorensen1999quantum} works in the continuous interaction regime while Milburn gate~\cite{milburn1999simulating} is in the pulsed interaction regime. In both cases, the collective vibrational motion of the ions are only virtually excited, thus relaxing the requirement of vibrational ground state cooling. However, this is true only if the mechanical phase space trajectory forms a closed loop. It has been pointed out in Ref.~\cite{sorensen2000entanglement} that, in the weak-field coupling limit, the internal state of the ions is independent of the state of the collective vibrational motion for any interaction time. In the mechanical phase space picture, this can be understood as following. In the limit of infinitesimal circle, we can hardly distinguish between open and closed loops. The total enclosed area is kept finite by traversing the phase space infinitely many times. Three natural questions follow. Given that, in S\o rensen-M\o lmer gate, the interaction time is subject to an error, which cannot be directly measured and compensated, the mechanical phase space trajectory is no longer closed. In this case, will the transformation of reducing the size of phase space loop together with traversing the phase space multiple times improve the gate performance? Will the same method help improve the performance of Milburn gate? How are the behaviors of the two gates connected with each other?

In this paper, we explicitly study these questions mentioned above. We consider S\o rensen-M\o lmer gate as a continuous interaction model between a target mode and an auxiliary mechanical oscillator mode, in the form described in Ref.~\cite{sorensen2000entanglement} but without restricting it to a trapped ion system. We consider Milburn gate as a series of pulsed interactions between a target mode and an auxiliary mechanical oscillator mode. We illustrate geometric interpretations of the two gates in the mechanical phase space. We explicitly show how S\o rensen-M\o lmer gate is equivalent to the continuous limit of Milburn gate. We then consider imperfect gates, with relative time error for S\o rensen-M\o lmer gate and phase error for Milburn gate. We study the transformation of decreasing the size of the loop together with traversing the phase space multiple times. The purity of both gates increases, but the fidelity of the two gates behave differently. We show that the difference in the fidelity is because the continuous limit of the pulsed scheme involves an error-dependent interaction strength. Finally we illustrate the analysis using an optomechanical system as an example~\cite{armata2016quantum,ma2020optical,vanner2011pulsed}, including the dissipation of the mechanical oscillator. We derive analytical solutions of the system in the pulsed interaction regime, and further obtain the results in the continuous interaction regime by taking the continuous limit without solving differential equations. The results enrich our understanding of the unification of the two interaction regimes.

\section{The original gates}

We briefly recapitulate the original S\o rensen-M\o lmer gate and Milburn gate in a trapped ion system. S\o rensen-M\o lmer gate~\cite{sorensen2000entanglement} uses bichromatic laser fields to excite the ions, which are slightly detuned from the upper and lower sidebands of a collective center-of-mass vibrational mode of the ions. In the Lamb-Dicke regime, the interaction Hamiltonian corresponds to a unitary time evolution operator in the form of $\hat{U}(t)=\exp[-iA(t)\hat{J}_y^2]\exp[-iF(t)\hat{J}_y\hat{x}]\exp[-iG(t)\hat{J}_y\hat{p}]$, where $\hat{J}_y$ is the collective spin operator, $\hat{x}\ (\hat{p})$ is the position (momentum) operator of the collective vibrational mode. The pair $(F(t),G(t))$ traverses a circular loop in the mechanical phase space. At times when the loop is closed, only the first term in $\hat{U}(t)$ is left, and $A(t)$ equals to the enclosed area of the circle. The gate is thus independent of the state of vibrational mode. Milburn gate~\cite{milburn1999simulating} uses bichromatic laser fields on resonance with the two sidebands of the vibrational motion of the ions. By properly choosing the phases of a sequence of laser pulses, the gate becomes $\hat{U}=\exp(i\kappa_x\hat{x}\hat{J}_z)\exp(i\kappa_p\hat{p}\hat{J}_z)\exp(-i\kappa_x\hat{x}\hat{J}_z)\exp(-i\kappa_p\hat{p}\hat{J}_z)=\exp(-i\kappa_x\kappa_p\hat{J}_z^2)$. It corresponds to a closed rectangle in the mechanical phase space. Similar to S\o rensen-M\o lmer gate, the gate is independent of the vibrational motion.

\section{Generalisation of Milburn gate}

The original Milburn gate~\cite{milburn1999simulating} describes a series of four pulsed interactions, forming a closed rectangle in the mechanical phase space. Here we generalise it to an arbitrary number of pulses, without the requirement of closing the mechanical phase space trajectory. We consider the case of equal interaction strength for each pulse and equal phase angle difference between adjacent pulses,
\begin{align}
\hat{U}_p=&\exp[i\lambda\hat{O}\frac{1}{\sqrt{2}}(\hat{b}+\hat{b}^{\dagger})]\times\exp[i\lambda\hat{O}\frac{1}{\sqrt{2}}(\hat{b}e^{i\theta}+\hat{b}^{\dagger}e^{-i\theta})]\nonumber\\
&\times\exp[i\lambda\hat{O}\frac{1}{\sqrt{2}}(\hat{b}e^{i2\theta}+\hat{b}^{\dagger}e^{-i2\theta})]\cdots\nonumber\\
&\times\exp[i\lambda\hat{O}\frac{1}{\sqrt{2}}(\hat{b}e^{i(N_p-1)\theta}+\hat{b}^{\dagger}e^{-i(N_p-1)\theta})]\nonumber\\
=&\exp[i\hat{O}(c_1 \hat{x}_m-c_2\hat{p}_m)]\times\exp[i\hat{O}^2c_3],
 \label{eq:pulse}
\end{align}
where $\hat{O}$ is an operator for the target mode, $N_p$ is the number of pulses, $\lambda$ is the dimensionless interaction strength, $\hat{b}\ (\hat{b}^{\dagger})$ is the annihilation (creation) operator on the auxiliary mechanical mode, $\theta$ is the phase angle increment. The second equality is a closed form expression as a result of the Baker-Campbell-Hausdorff formula~\cite{sakurai2014modern}, where $c_1=\sum_{n=0}^{N_p-1}\cos(n\theta)$,  $c_2=\sum_{n=0}^{N_p-1}\sin(n\theta)$, $\hat{x}_m$ and $\hat{p}_m$ are dimensionless position and momentum operator of the mechanical mode, $\hat{x}_m=(\hat{b}+\hat{b}^{\dagger})/\sqrt{2}$, $\hat{p}_m=i(\hat{b}^{\dagger}-\hat{b})/\sqrt{2}$. The explicit expressions of $c_1$, $c_2$ and $c_3$ are shown in Appendix~\ref{app:0}.

\begin{figure}[t]
\centering
\includegraphics[width=0.35\textwidth]{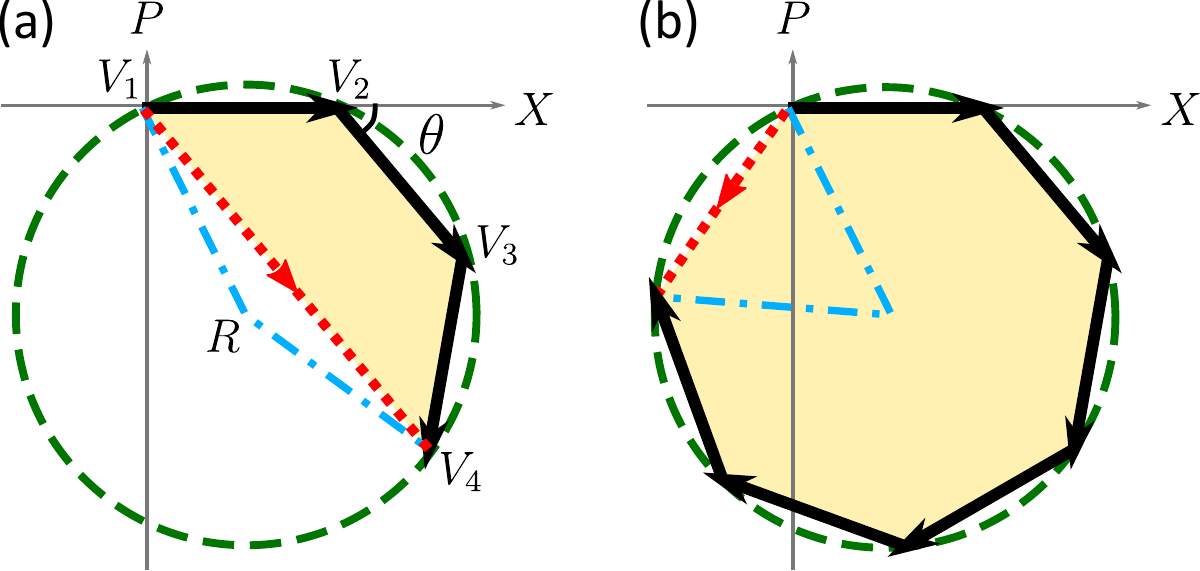}
\caption{Geometric explanation of Milburn gate in the mechanical phase space. $X(-P)$ component of the red dotted vector represents $c_1(c_2)$, the area of the yellow shaded region represents $c_3$. (a) $\theta_{\mathrm{net}}<\pi$. (b) $\theta_{\mathrm{net}}>\pi$.}\label{fig:1}
\end{figure}

We associate geometric meanings to the coefficients $c_1$, $c_2$ and $c_3$, as shown in Fig.~\ref{fig:1}. Each pulse in Eq.~\eqref{eq:pulse} is depicted as a thick black vector $\overrightarrow{V_iV_{i+1}}$ with length $\lambda$. The phase angle increment $\theta$ is the angle between two adjacent vectors. All the start and end points of the vectors lie on a circle, with its centre labelled as $R$. The red dotted vector $\overrightarrow{V_1V_{N_p+1}}$ connects the start point of the first vector and the end point of the last vector. Its component in $X$ axis is $c_1$, and its component in $-P$ axis is $c_2$. $c_3$ is the difference between two areas. The first is $N_p$ times the area of the triangle $\bigtriangleup V_iV_{i+1}R$. The second is the signed-area of the triangle $\bigtriangleup V_1RV_{N_p+1}$. We define the net swept angle $\theta_{\mathrm{net}}$ as $\theta_{\mathrm{net}}=N_p\theta-2M\pi$, with $M$ as a non-negative integer so that $0\leq\theta_{\mathrm{net}}<2\pi$.  If $\theta_{\mathrm{net}}<(>)\pi$, the positive (negative) sign is taken. These two situations are plotted in Fig.~\ref{fig:1}(a) and (b), respectively, for the simple case of $M=0$. Note that, if the mechanical phase space trajectory displayed here by the thick black vectors are rotated around the coordinate centre by $90^{\circ}$ counterclockwise, it becomes the same as phase space trajectory defined by unitary transformation of operators, $\hat{U}^{\dagger}\hat{x}_m\hat{U}=\exp[-i(c_1\hat{x}_m-c_2\hat{p}_m)]\hat{x}_m\exp[i(c_1\hat{x}_m-c_2\hat{p}_m)]=\hat{x}_m+c_2$, and $\hat{U}^{\dagger}\hat{p}_m\hat{U}=\exp[-i(c_1\hat{x}_m-c_2\hat{p}_m)]\hat{p}_m\exp[i(c_1\hat{x}_m-c_2\hat{p}_m)]=\hat{p}_m+c_1$.

\section{S\o rensen-M\o lmer gate as the continuous limit of Milburn gate}~\label{sec:SMandG}

The continuous regime can be derived from taking the continuous limit of the pulsed regime. To be specific, we define the rescaled interaction strength $k=\lambda/\sqrt{2}\theta$, and take the limits $\theta\rightarrow0$, $N_p\rightarrow\infty$, while keeping the product $N_p\theta=\phi$ as a constant, with the angle $\phi$ proportional to the interaction time, $\phi=\omega_mt$, where $\omega_m$ is the frequency of the mechanical mode. The resulting gate becomes
\begin{subequations}\label{eq:s}
\begin{align}
&\hat{U}_c=\exp[i\hat{O}(d_1 \hat{x}_m-d_2\hat{p}_m)]\times\exp[i\hat{O}^2d_3],\\
&d_1=\sqrt{2}k\sin\omega_mt,\\
&d_2=\sqrt{2}k(1-\cos\omega_mt),\\
&d_3=k^2(\omega_mt-\sin\omega_mt).
\end{align}
\end{subequations}
This is in the form of continuous interaction given by S\o rensen-M\o lmer gate~\cite{sorensen2000entanglement}. Coefficients $d_1$, $d_2$ and $d_3$ also have geometric meanings, as shown in Fig.~\ref{fig:2}. The continuous interaction is represented by the thick black arc with start point $L_1$ at the coordinate centre and end point $L_2$ on a circle with radius $\sqrt{2}k$ whose centre $R$ is on the $-P$ axis. The angle swept by $|L_2R|$ from $|L_1R|$ is $\phi$. The $X(-P)$ component of the red dotted vector $\overrightarrow{L_1L_2}$ equals to $d_1(d_2)$. $d_3$ is given by the difference between two areas. Suppose $\phi=2\pi M+\phi_{\mathrm{net}}$, where $M$ is a non-negative integer and $0\leq \phi_{\mathrm{net}}<2\pi$. The first area is the area of $M$ circles plus the area of the circular sector formed by the arc $L_1L_2$ and the two radii $|L_1R|$ and $|L_2R|$, with central angle $\phi_{\mathrm{net}}$. The second area is the signed-area of the triangle $\bigtriangleup L_1RL_2$, taking a positive (negative) sign for $\phi_{\mathrm{net}}<(>)\pi$. Fig.~\ref{fig:2} shows the two cases for $M=0$.


\section{Including relative error}

Both S\o rensen-M\o lmer gate and Milburn gate have the property that, if the mechanical phase space trajectory forms a closed loop, the gate operator becomes independent of the mechanical oscillator mode, thus removing the necessity of mechanical cooling. However, any error in parameters of the gate will violate this condition, entangling the target mode and the auxiliary mechanical mode. As shown in Fig.~\ref{fig:1} and~\ref{fig:2}, the amount of entanglement is quantified by the length of the red dotted vector, which is bounded to the green dashed circle. A straightforward strategy to suppress the amount of entanglement is to reduce the interaction strength and in compensation traverse the mechanical phase space multiple times, an idea originating from the weak-field coupling regime in Ref.~\cite{sorensen2000entanglement}. But it is not straightforward how the gate fidelity is affected by this transformation. In this section, we first study  S\o rensen-M\o lmer gate when there is relative error in controlling the interaction time. We then study Milburn gate when there is error in the phase angle increment $\theta$. Finally we discuss the way of unifying the two results in one mathematical framework.

\subsection{Relative error in interaction time for S\o rensen-M\o lmer gate}\label{sec:s}

\begin{figure}[t]
\centering
\includegraphics[width=0.35\textwidth]{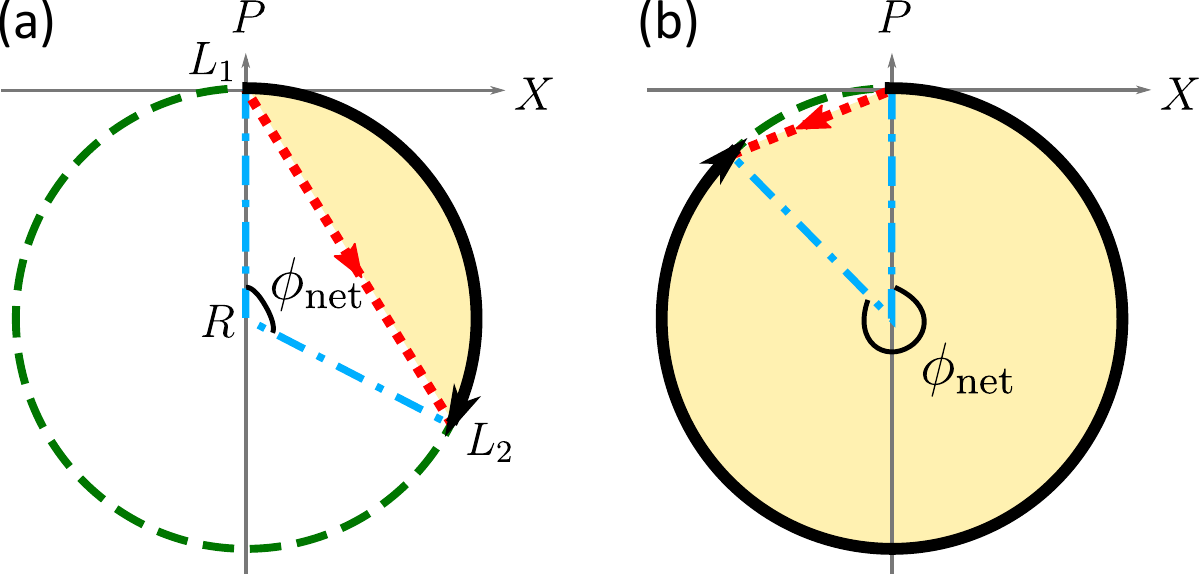}
\caption{Geometric explanation of S\o rensen-M\o lmer gate in the mechanical phase space. $X(-P)$ component of the red dotted vector represents $d_1(d_2)$, the area of the yellow shaded region represents $d_3$. (a) $\phi_{\mathrm{net}}<\pi$. (b) $\phi_{\mathrm{net}}>\pi$.}\label{fig:2}
\end{figure}

It is straightforward to deduce from the S\o rensen-M\o lmer gate expression Eq.~\eqref{eq:s} that, no entanglement between the target mode and the mechanical mode is generated if $\omega_mt=2K\pi$ with $K$ a positive integer. For simplicity we assume $K=1$. Generalisation to other values of $K$ is straightforward. Consider the case that the interaction time $t'$ cannot be controlled precisely, so that $t'=(1+\eta)2\pi/\omega_m$. $\eta$ characterises the relative error of the interaction time, $|\eta|\ll1$. The gate can be decomposed into $\hat{U}_{c,N=1}(\eta)=\hat{V}_{m,N=1}\hat{V}_{O,N=1}\hat{U}_{c,T}$, where $\hat{V}_{m,N=1}$ is the error gate induced by entanglement with the mechanical mode, $\hat{V}_{O,N=1}$ is the error gate induced by an effective additional self-interaction, and $\hat{U}_{c,T}$ is the target gate. The explicit expressions are listed in Appendix~\ref{app:0}, and also correspond to Eq.~\eqref{eq:SM_for_N} below taking $N=1$. For $|\eta|\ll1$, $\hat{V}_{O,N=1}$ is approximately the identity operator. This can be understood from Fig.~\ref{fig:2}. For simplicity suppose $\eta>0$, then $\phi_{\mathrm{net}}=2\pi\eta$ [Fig.~\ref{fig:2}(a)]. The yellow shaded area represents the exponential of $\hat{V}_{O,N=1}$, which is proportional to $O(\eta^2)$. 

Consider reducing the interaction strength by a factor of $N$, $k\rightarrow k/N$. In order to reproduce the same target gate, the interaction time needs to increase by a factor of $N^2$, $t'\rightarrow N^2t'$. The gate becomes
\begin{subequations}\label{eq:SM_for_N}
\begin{align}
&\hat{U}_{c,N}(\eta)=\hat{V}_{m,N}\hat{V}_{O,N}\hat{U}_{c,T},\\
&\hat{V}_{m,N}=e^{i\sqrt{2}\frac{k}{N}\hat{O}[\sin(\eta2\pi N^2)\hat{x}_m-(1-\cos(\eta2\pi N^2))\hat{p}_m]},\\
&\hat{V}_{O,N}=\exp\{ik^2\hat{O}^2[\eta2\pi-\frac{\sin(\eta2\pi N^2)}{N^2}]\},\\
&\hat{U}_{c,T}=\exp(ik^2\hat{O}^22\pi).
\end{align}
\end{subequations}
with $\eta$ the same relative error of the interaction time as before. The target mode gets completely disentangled with the mechanical mode for all values of $|\eta|\ll1$ if the limit $N\rightarrow\infty$ is taken, where the gate turns out to be $\hat{U}_{c,N\rightarrow\infty}(\eta)=\hat{V}_{O,N\rightarrow\infty}\hat{U}_{c,T}$ with $\hat{V}_{O,N\rightarrow\infty}=\exp(ik^2\hat{O}^2\eta 2\pi)$. Note that $\hat{V}_{O,N\rightarrow\infty}$ is not an identity operator, indicating a finite error in the self-interaction.

Whether reducing the interaction strength together with increasing the interaction time improves the gate fidelity depends on the comparison between the impact of $\hat{V}_{m,N=1}$ and $\hat{V}_{O,N\rightarrow\infty}$. For instance, if the mechanical mode is initially in a high temperature thermal state, $\hat{V}_{m,N=1}$ dominates $\hat{V}_{O,N\rightarrow\infty}$. The transformation improves the fidelity. In contrast, the purity of the target mode always increases after the transformation, regardless of the relative impact of $\hat{V}_{m,N=1}$ and $\hat{V}_{O,N\rightarrow\infty}$.

\subsection{Error in phase angle increment for Milburn gate}\label{sec:m}

Milburn gate Eq.~\eqref{eq:pulse} forms a regular polygon in the mechanical phase space if $\theta=2\pi/N_p$, so that the target mode disentangles with the mechanical mode. The target gate is therefore
\begin{equation}
\hat{U}_{p,T}=\exp[i\lambda^2\hat{O}^2\frac{N_p}{4}\cot(\frac{\pi}{N_p})].
\end{equation}
Note that for simplicity we have assumed that the mechanical phase space is traversed once.

We consider an error in controlling the phase angle increment $\theta$. This is also an error in controlling time if the phase angle increment is implemented by leaving the mechanical oscillator mode to evolve freely for a certain amount of time, as in a pulsed optomechanical system which we will discuss later~\cite{vanner2011pulsed}. Suppose the phase angle increment is $\theta'=(1+\xi)2\pi/N_p$, where $\xi$ is the relative error satisfying $|\xi|\ll1$. The mechanical phase space trajectory becomes open, indicating entanglement between the two modes. Similar to the idea in S\o rensen-M\o lmer gate, the entanglement decreases after the reducing the interaction strength $\lambda\rightarrow\lambda/N$ together with increasing the number of pulses $N_p\rightarrow N^2N_p$, for $N$ as a positive integer. Note that $\theta'$ is not changed by this transformation. The mechanical phase space is traversed multiple times and the gate becomes
\begin{subequations}\label{eq:MilburnForN}
\begin{align}
&\hat{U}_{p,N}(\xi)=\exp(i\hat{\psi}_{m,N})\exp(i\hat{\psi}_{O,N})\hat{U}_{p,T},\\
&\hat{\psi}_{m,N}=\frac{\lambda}{N}\hat{O}\{[\frac{1}{2}+\frac{1}{2}\cos(N^2\xi2\pi-\frac{2\pi}{N_p}-\xi\frac{2\pi}{N_p})\\\nonumber
&\ \ +\frac{1}{2}\sin(N^2\xi2\pi-\frac{2\pi}{N_p}-\xi\frac{2\pi}{N_p})\cot(\frac{\pi}{N_p}+\xi\frac{\pi}{N_p})]\hat{x}_m\\\nonumber
&\ \ -[\frac{1}{2}\cot(\frac{\pi}{N_p}+\xi\frac{\pi}{N_p})+\frac{1}{2}\sin(N^2\xi2\pi-\frac{2\pi}{N_p}-\xi\frac{2\pi}{N_p})\\\nonumber
&\ \ -\frac{1}{2}\cos(N^2\xi2\pi-\frac{2\pi}{N_p}-\xi\frac{2\pi}{N_p})\cot(\frac{\pi}{N_p}+\xi\frac{\pi}{N_p})]\hat{p}_m\},\\
&\hat{\psi}_{O,N}=\lambda^2\hat{O}^2\{\frac{N_p}{4}[\cot(\frac{\pi}{N_p}+\xi\frac{\pi}{N_p})-\cot(\frac{\pi}{N_p})]\\\nonumber
&\ \ -\frac{\sin(N^2\xi2\pi)}{8N^2\sin^2(\frac{\pi}{N_p}+\xi\frac{\pi}{N_p})}\}.
\end{align}
\end{subequations}

To check whether the transformation of reducing the interaction strength together with increasing the number of pulses improves the gate performance, we compare the error gates of the original gate $(N=1)$ and the limit $N\rightarrow\infty$. $\exp(i\hat{\psi}_{m,N\rightarrow\infty})$ is an identity operator, but $\exp(i\hat{\psi}_{m,N=1})$ is not. Therefore the purity of the target mode always increases as $N$ increases. Taylor expansion of $\hat{\psi}_{O,N=1}$ in terms of $\xi$ keeping up to the linear term leads to
\begin{equation}\label{eq:error1}
\hat{\psi}_{O,N=1}\approx-\lambda^2\hat{O}^2\frac{\pi\xi}{2\sin^2(\frac{\pi}{N_p})}.
\end{equation}
Note this is different from the case of S\o rensen-M\o lmer gate, where the effective additional self-interaction is zero to the first order of the relative error. It is because here both the central angle and the radius of the circle change with the phase angle increment $\theta'$ [Fig.~\ref{fig:1}(a)]. Similarly, Taylor expansion of $\hat{\psi}_{O,N\rightarrow\infty}$ gives
\begin{equation}\label{eq:error2}
\hat{\psi}_{O,N\rightarrow\infty}\approx-\lambda^2\hat{O}^2\frac{\pi\xi}{4\sin^2(\frac{\pi}{N_p})}.
\end{equation}
The effective additional self-interaction is thus also reduced by the transformation. As a result, the fidelity of Milburn gate is expected to improve under this transformation.

\subsection{Unifying the two schemes}

A natural question arises here. Given that S\o rensen-M\o lmer gate is the continuous limit of Milburn gate, as demonstrated in Sec.~\ref{sec:SMandG}, why do the fidelities of the two gates behave differently in the presence of relative error and upon the transformation of decreasing the interaction strength together with traversing the phase space multiple times?

After examining the continuous limit of Milburn gate (see the paragraph above Eq.~\eqref{eq:s}), we point out that, Eq.~\eqref{eq:SM_for_N} is indeed the continuous limit of Eq.~\eqref{eq:MilburnForN}. To be specific, if we replace the interaction strength $\lambda$ in Eq.~\eqref{eq:MilburnForN} with
\begin{equation}\label{eq:int_strength}
    \lambda=\sqrt{2}k(1+\xi)\frac{2\pi}{N_p},
\end{equation}
and take the limit $N_p\rightarrow\infty$, we get Eq.~\eqref{eq:SM_for_N} after renaming $\xi$ to $\eta$. Note that in Eq.~\eqref{eq:int_strength}, $\lambda$ explicitly depends on the relative error $\xi$. This dependence results in the difference between $\exp(i\hat{\psi}_{O,N=1})$ and $\hat{V}_{O,N=1}$, and the difference between $\exp(i\hat{\psi}_{O,N\rightarrow\infty})$ and $\hat{V}_{O,N\rightarrow\infty}$. Indeed, if we insert Eq.~\eqref{eq:int_strength} into the error estimations for Milburn gate, Eqs.~\eqref{eq:error1} and~\eqref{eq:error2}, and expand to first order in $\xi$, we recover the error estimation for S\o rensen-M\o lmer gate. In Appendix~\ref{app:0}, we use a diagram to show how the expressions in the two regimes are related to each other via equalities and limits.

Equation~\eqref{eq:int_strength}, together with the original rescaling relation $\lambda=\sqrt{2}k\theta$ above Eq.~\eqref{eq:s}, connects the pulsed interaction regime with the continuous interaction regime. This unification will be further illustrated with an example in optomechanics. 

\section{Optomechanical model as an example}

\begin{figure}[t]
\centering
\includegraphics[width=0.48\textwidth]{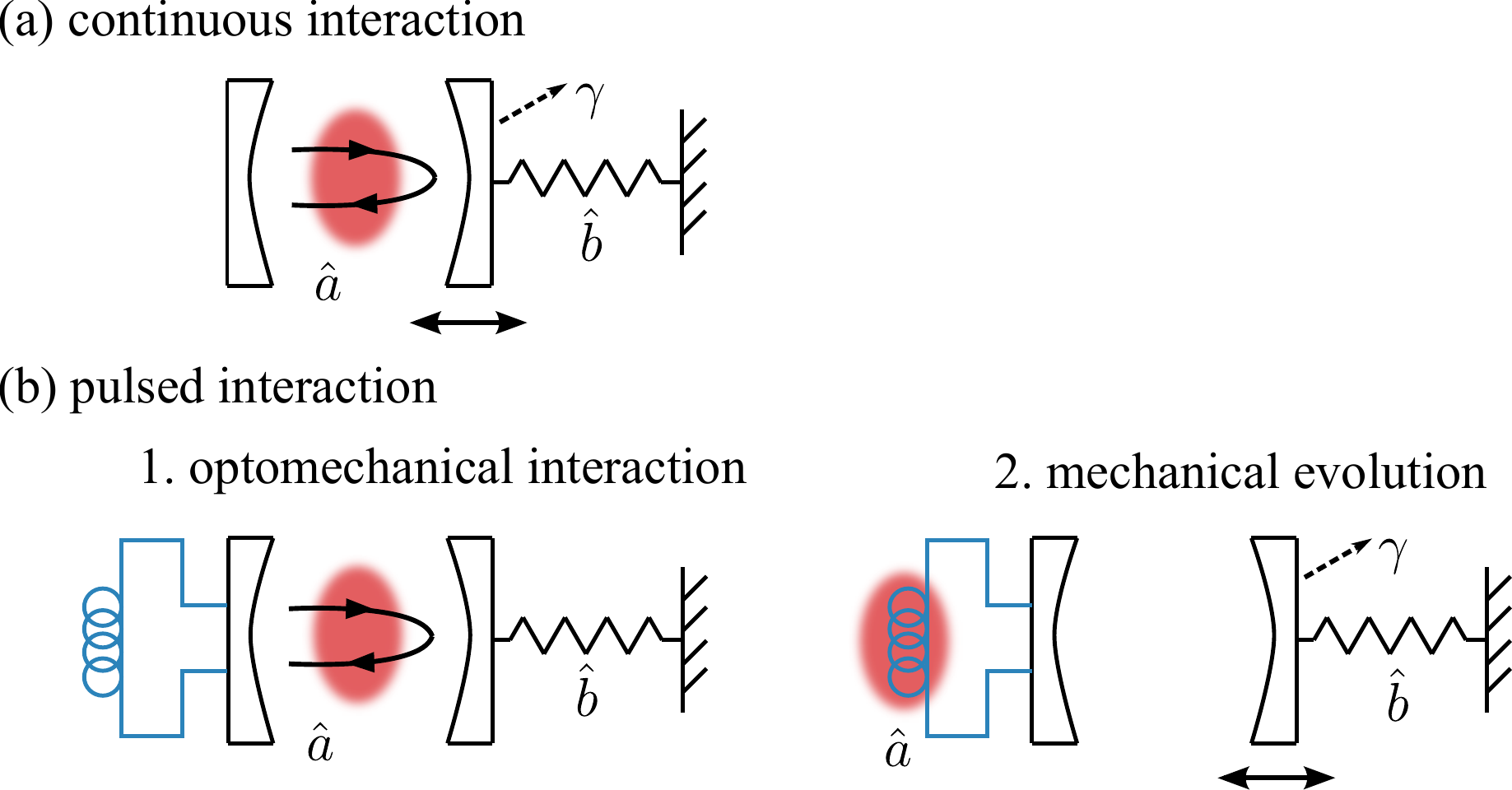}
\caption{The optomechanical model is illustrated as a Fabry-P\'{e}rot cavity here. The optical field circulates inside the cavity (red circle, described by the annihilation operator $\hat{a}$), one end-mirror of which is movable and modelled as a mechanical oscillator (annihilation operator $\hat{b}$). The optical field exerts a radiation pressure on the mechanical oscillator, while the mechanical oscillator modifies the optical path of the field. (a) In the continuous interaction regime, the optical field stays in the cavity for multiple mechanical periods. The optomechanical interaction happens simultaneously with the free evolution of the mechanical oscillator (double-headed arrow) and the possible mechanical dissipation (dashed arrow, $\gamma$ as the dissipation rate). (b) In the pulsed interaction regime, there are two separate steps that are repeated periodically. The first step is the optomechanical interaction. It is on a much shorter time scale than the mechanical period, therefore in this step both the mechanical oscillator free evolution and dissipation are neglected. In the second step, the optical field exits the cavity to enter a delay line (illustrated as the blue circuit). The mechanical oscillator, at the same time, evolves freely for a fraction of the mechanical period, together with the possible mechanical dissipation. The two steps are then repeated.}\label{fig:om}
\end{figure}

We have already discussed about the general abstract form of Milburn gate and S\o rensen-M\o lmer gate. We focused on the presence of a relative error in the gate implementation. For S\o rensen-M\o lmer gate, we considered the error in the gate implementation time, while for Milburn gate, we considered the error in the phase angle increment. We analysed the performance of the gates if we decrease the interaction strength and traverse the mechanical phase space multiple times. The latter refers to increasing the interaction time for S\o rensen-M\o lmer gate and increasing the number of pulses for Milburn gate. In this section, we apply our results to an optomechanical system (see Fig.~\ref{fig:om}), where S\o rensen-M\o lmer gate corresponds to the continuous interaction regime, and Milburn gate corresponds to the pulsed interaction regime. We will also show that, the phase angle increment in Milburn gate is also related to an evolution time. Additionally, we take the dissipation of the mechanical oscillator into account. By deriving the analytical solutions, we further comment on the relations between Milburn gate and S\o rensen-M\o lmer gate.

\subsection{Unitary dynamics}

We consider an example of optomechanical system to analytically quantify the performance of the gates under the transformation described above. An optomechanical system consists of a 
Fabry-P\'{e}rot cavity with a movable mirror~\cite{aspelmeyer2014cavity}. Light field inside the cavity interacts with the mechanical oscillation of the movable mirror via radiation pressure. The Hamiltonian in the frame rotating with the field frequency is~\cite{armata2016quantum,ma2020optical}
\begin{equation}\label{eq:OM_hamiltonian}
\hat{H}=\hbar\omega_m\hat{b}^{\dagger}\hat{b}-\hbar g_0\hat{a}^{\dagger}\hat{a}\frac{\hat{b}^{\dagger}+\hat{b}}{\sqrt{2}},
\end{equation}
where $g_0$ is the optomechanical interaction strength, $\hat{a}(\hat{a}^{\dagger})$ is the annihilation (creation) operator of the field.

If the cavity photon decay rate $\kappa$ is small, the field is kept in the cavity for multiple mechanical periods, the system is in the continuous interaction regime, see Fig.~\ref{fig:om}(a). The time evolution operator is thus calculated to be~\cite{mancini1997ponderomotive}
\begin{equation}
\begin{split}
&\hat{U}^{(\mathrm{OM})}_{c,1}(t)=e^{-i\hat{H}t/\hbar}\\
&=\exp\{i\frac{g_0}{\omega_m}\hat{a}^{\dagger}\hat{a}[\sin(\omega_mt)\hat{x}_m-(1-\cos(\omega_mt))\hat{p}_m]\}\\
&\ \ \times\exp\{i\frac{g_0^2}{2\omega_m^2}(\hat{a}^{\dagger}\hat{a})^2[\omega_mt-\sin(\omega_mt)]\}\\
&\ \ \times\exp(-i\omega_mt\hat{b}^{\dagger}\hat{b}).
\end{split}
\end{equation}
The last term induces a uniform rotation in the mechanical phase space. It has no effect if the mechanical mode is initially in a thermal state, which is the case we will focus on. $\hat{U}^{(\mathrm{OM})}_{c,1}(t)$ is in the form of S\o rensen-M\o lmer gate with dimensionless interaction strength $k=g_0/\sqrt{2}\omega_m$ and $\hat{O}=\hat{a}^{\dagger}\hat{a}$. 

As in Sec.~\ref{sec:s}, suppose that the target gate corresponds to the interaction time $t=2\pi/\omega_m$ and there is a relative error in the actual interaction time, so that  $t'=(1+\eta)t$. We investigate the change in gate performance as we reduce the interaction strength $k\rightarrow k/N$ and increase the interaction time $t'\rightarrow N^2t'$. The initial state of the field is assumed to be a coherent state $|\alpha\rangle_f$. The mechanical oscillator is initialised to a thermal state with mean phonon number $n_{\mathrm{th}}$. Following the discussion in Sec.~\ref{sec:s}, the purity of the final field state is improved by taking a large value of $N$. The fidelity depends on the comparison between $\hat{V}_{m,N=1}$ and $\hat{V}_{O,N\rightarrow\infty}$. We consider the change of the mean value of the field quadrature operator induced by these two error gates, which is the lowest order contribution. Suppose $\alpha$ is real without loss of generality. For $\hat{V}_{m,N=1}$,
\begin{equation}
\begin{split}
&\langle \hat{V}_{m,N=1}^{\dagger} (\hat{a}e^{-i\varphi}+\hat{a}^{\dagger}e^{i\varphi}) \hat{V}_{m,N=1} \rangle\\
&=2\alpha\cos\varphi \exp\{-(2n_{\mathrm{th}}+1)k^2[1-\cos(\eta 2 \pi)]\},\\
\end{split}
\end{equation}
inducing a decay of the field quadrature amplitude. For $\hat{V}_{O,N\rightarrow\infty}$, 
\begin{equation}
\begin{split}
&\langle \hat{V}_{O,N\rightarrow\infty}^{\dagger} (\hat{a}e^{-i\varphi}+\hat{a}^{\dagger}e^{i\varphi}) \hat{V}_{O,N\rightarrow\infty} \rangle\\
&=2\alpha e^{-\alpha^2[1-\cos(4\pi\eta k^2)]}\cos[\varphi-\alpha^2\sin(4\pi\eta k^2)-2\pi\eta k^2],\\
\end{split}
\end{equation}
resulting in both a decay in amplitude and a change in phase. For practical optomechanical parameters of solid state systems~\cite{aspelmeyer2014cavity}, the amplitude decays in both cases are negligibly small. It is thus expected that the change in the effective self-interaction induced by $\hat{V}_{O,N\rightarrow\infty}$ dominates the effect of $\hat{V}_{m,N=1}$. The transformation of reducing the interaction strength together with increasing the interaction time reduces the fidelity of the field state.

\begin{figure}[t]
\centering
\includegraphics[width=0.48\textwidth]{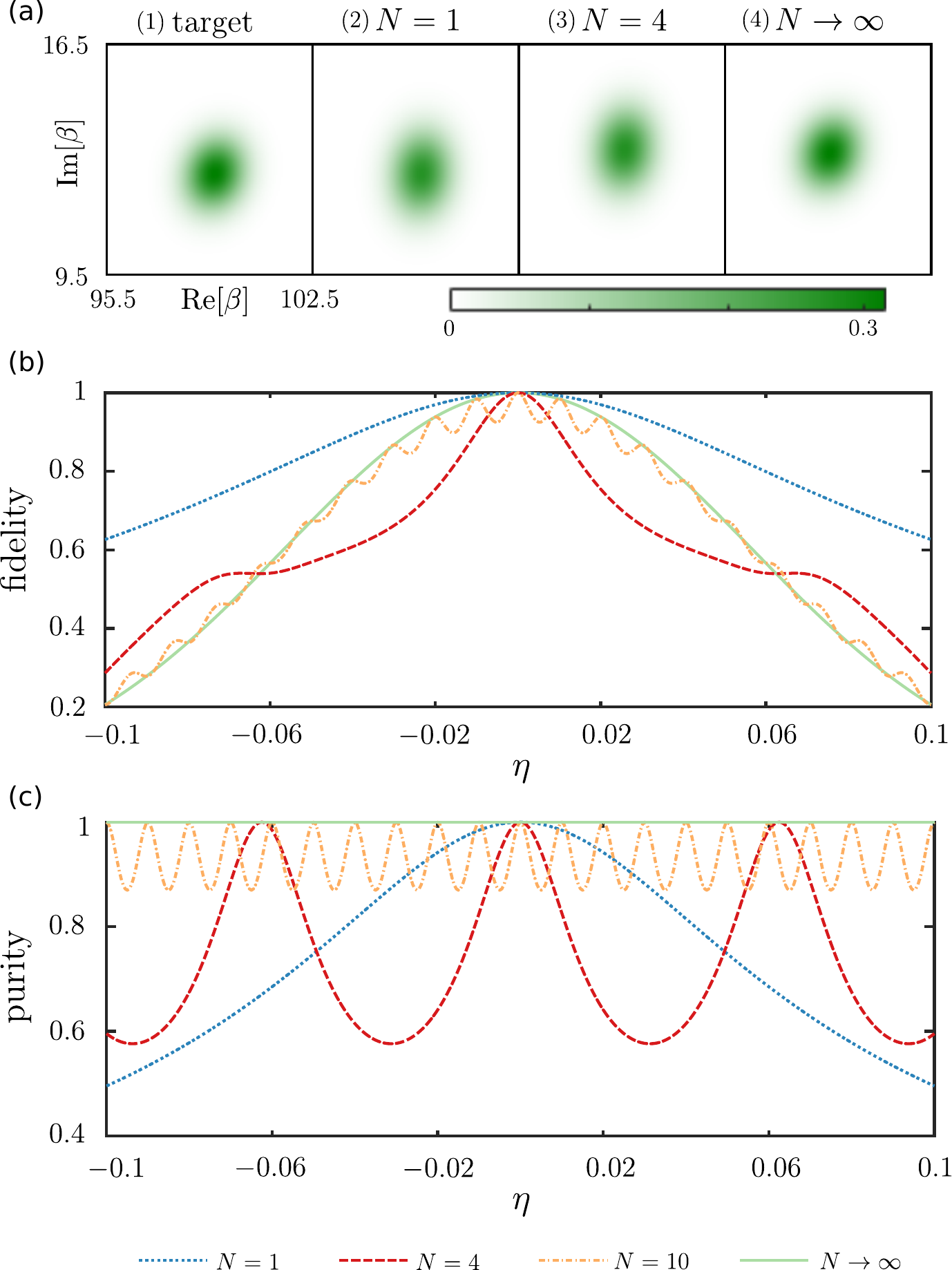}
\caption{Performance of S\o rensen-M\o lmer gate (continuous optomechanical interaction) for $\alpha=100$, $k=0.001$ and $n_{\mathrm{th}}=100$. (a) Q-function of the optical field for the target state [(1)] and states with relative error in the interaction time $\eta=0.05$, different values of $N$ [(2)-(4)]. (b) Fidelity of the final field state compared with the target state as a function of $\eta$, for different values of $N$. (c) Purity of the final field state as a function of $\eta$, for different values of $N$. For (b) and (c), blue dotted line is for $N=1$, red dashed line is for $N=4$, orange dot-dashed line is for $N=10$, green solid line is for $N\rightarrow\infty$.}\label{fig:3}
\end{figure}

\begin{figure}[t!]
\centering
\includegraphics[width=0.48\textwidth]{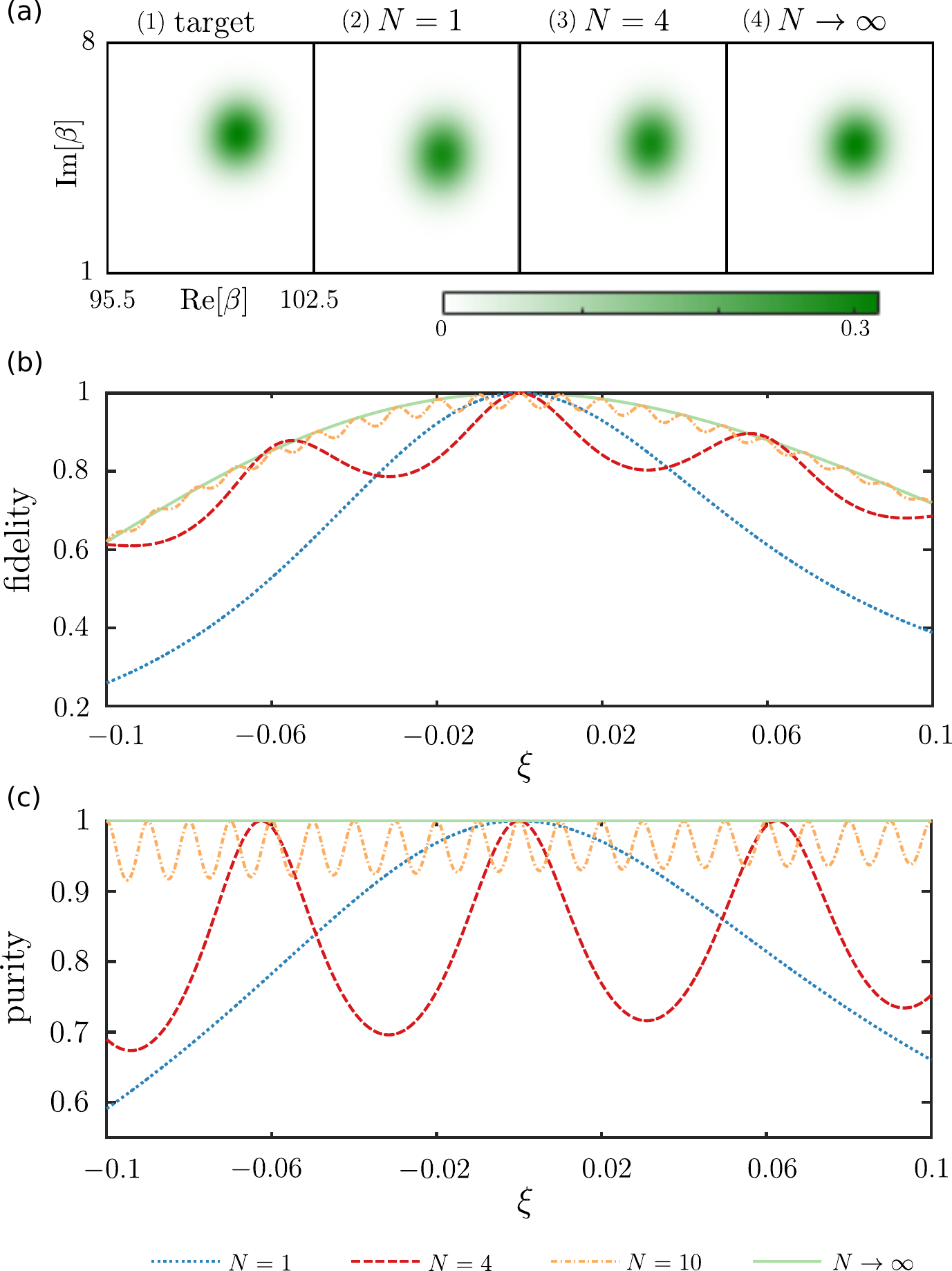}
\caption{Performance of Milburn gate (pulsed optomechanical interaction) for $\alpha=100$, $\lambda=0.001$, $N_p=6$, and $n_{\mathrm{th}}=100$. (a) Q-function of the field for the target state [(1)], and states with relative error in the phase angle increment $\xi=0.05$, different values of $N$ [(2)-(4)]. (b) Fidelity of the final field state as a function of $\xi$ for different $N$. (c) Purity of the final field state as a function of $\xi$ for different $N$. Legends are the same as in Fig.~\ref{fig:3}.}
\label{fig:4}
\end{figure}

Fig.~\ref{fig:3} shows an example of the gate performance under the transformation. We choose the amplitude of the coherent state $\alpha=100$, the dimensionless interaction strength $k=0.001$, and the thermal state with mean phonon number $n_{\mathrm{th}}=100$. Fig.~\ref{fig:3}(a) depicts the Q-function of the field state, $Q(\beta)={}_f\langle \beta|\rho_f |\beta\rangle_{f}/\pi$. Panel (1) is for the target state. It represents a self-Kerr interaction on a coherent state, inducing rotation and a small amount of squeezing~\cite{milburn1986quantum,ma2020optical}. Panel (2) is for the field state at time $t'=(1+\eta)2\pi/\omega_m$ with $\eta=0.05$. Entanglement with the thermal mechanical mode smears out the peak of the Q-function, while the centre of the peak remains unchanged. Panel (3) is for $t'=N^2(1+\eta)2\pi/\omega_m$ and dimensionless interaction strength $k/N$ with $\eta=0.05$ and $N=4$. The extra rotation induced by $\hat{V}_{O,N=4}$ is clear. Panel (4) is for the limit $N\rightarrow\infty$ with $\eta=0.05$. The field becomes disentangled with the mechanical oscillator, so the peak of the Q-function is concentrated. But the extra rotation means the overlap with the target state is smaller than that for $N=1$. Fig.~\ref{fig:3}(b) plots the fidelity of the final field state compared with the target state, as a function of the relative error in interaction time $\eta$, for $N=1$ (blue dotted), $N=4$ (red dashed), $N=10$ (orange dot-dashed) and $N\rightarrow\infty$ (green solid). The fidelity is defined as the overlap between the state of the field and the target state, which is
\begin{equation}
F_c={}_f\langle\alpha|e^{-i2\pi k^2(\hat{a}^{\dagger}\hat{a})^2}\rho_fe^{i2\pi k^2(\hat{a}^{\dagger}\hat{a})^2}|\alpha\rangle_{f}.
\end{equation}
As discussed before, the fidelity is reduced by the transformation due to the resulted error in the effective self-interaction. Note that $N=10$ already nicely approaches the limit $N\rightarrow\infty$. Fig.~\ref{fig:3}(c) plots the purity of the final field state as a function of $\eta$, for values of $N$ same as those in (b). The purity is defined as
\begin{equation}
P_c=\mathrm{Tr}(\rho_f^2).
\end{equation}
The purity is already largely improved for $N=10$, and will reach $1$ for any interaction time in the limit $N\rightarrow\infty$. The analytical expressions for the Q-function, fidelity and purity are shown in Appendix~\ref{app:unitary}.

\begin{figure}[t!]
\centering
\includegraphics[width=0.48\textwidth]{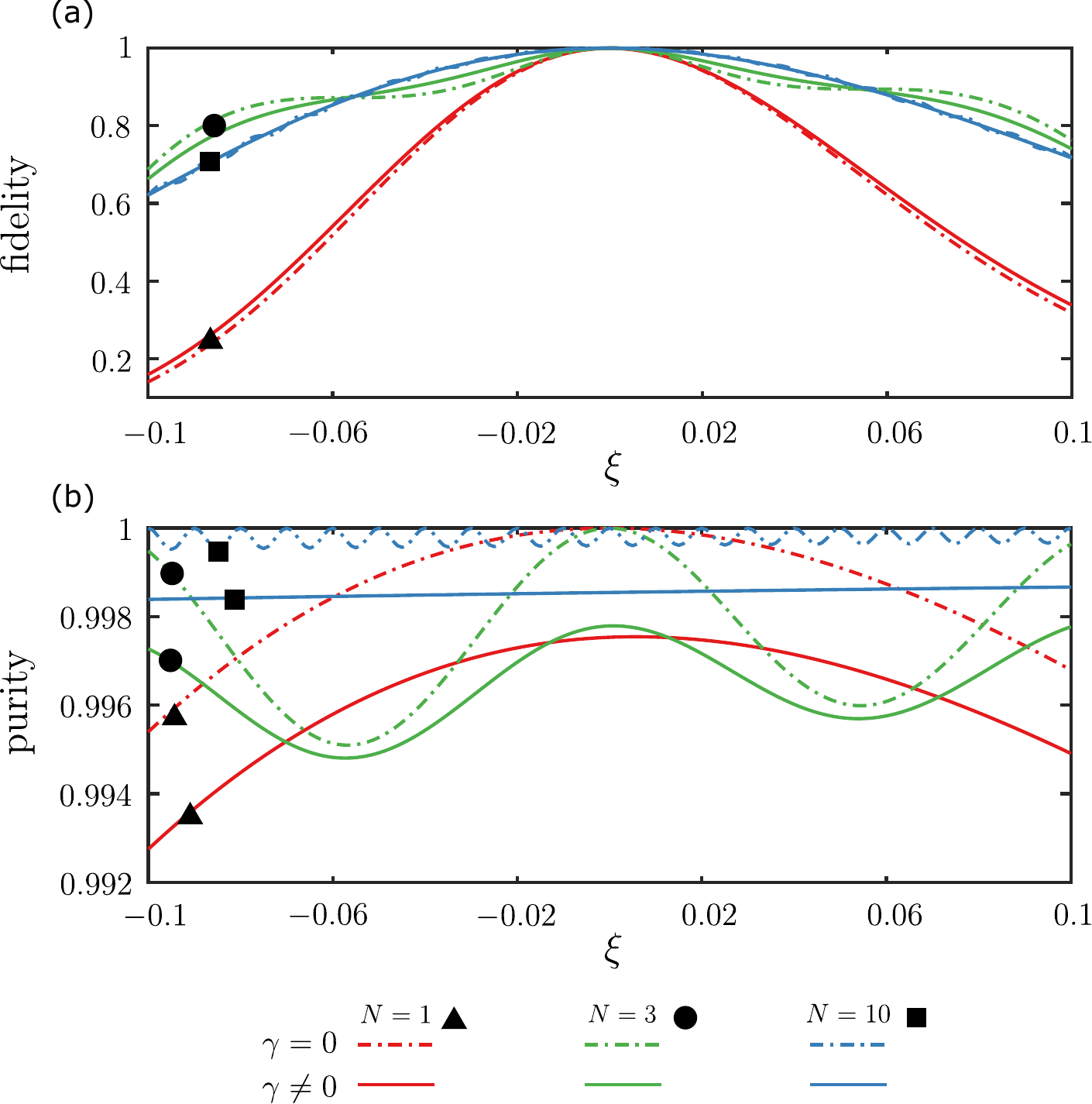}
\caption{Performance of Milburn gate (pulsed optomechanical interaction) including the dissipation of the mechanical oscillator described by the Lindblad master equation, for $\alpha=100$, $\lambda=0.001$, $N_p=6$, $n_{\mathrm{th}}=0$ and $\gamma/\omega_m=0.02$. (a) Fidelity of the final field state as a function of $\xi$ for different $N$. (b) Purity of the final field state as a function of $\xi$ for different $N$. Solid lines are for the cases with mechanical dissipation. For comparison, dot-dashed lines are for the cases without mechanical dissipation. The case of $N=1$ are plotted in red and marked with triangles. The case of $N=3$ are plotted in green and marked with circles. The case of $N=10$ are plotted in blue and marked with squares.}
\label{fig:m_dsp}
\end{figure}

\begin{figure}[t!]
\centering
\includegraphics[width=0.48\textwidth]{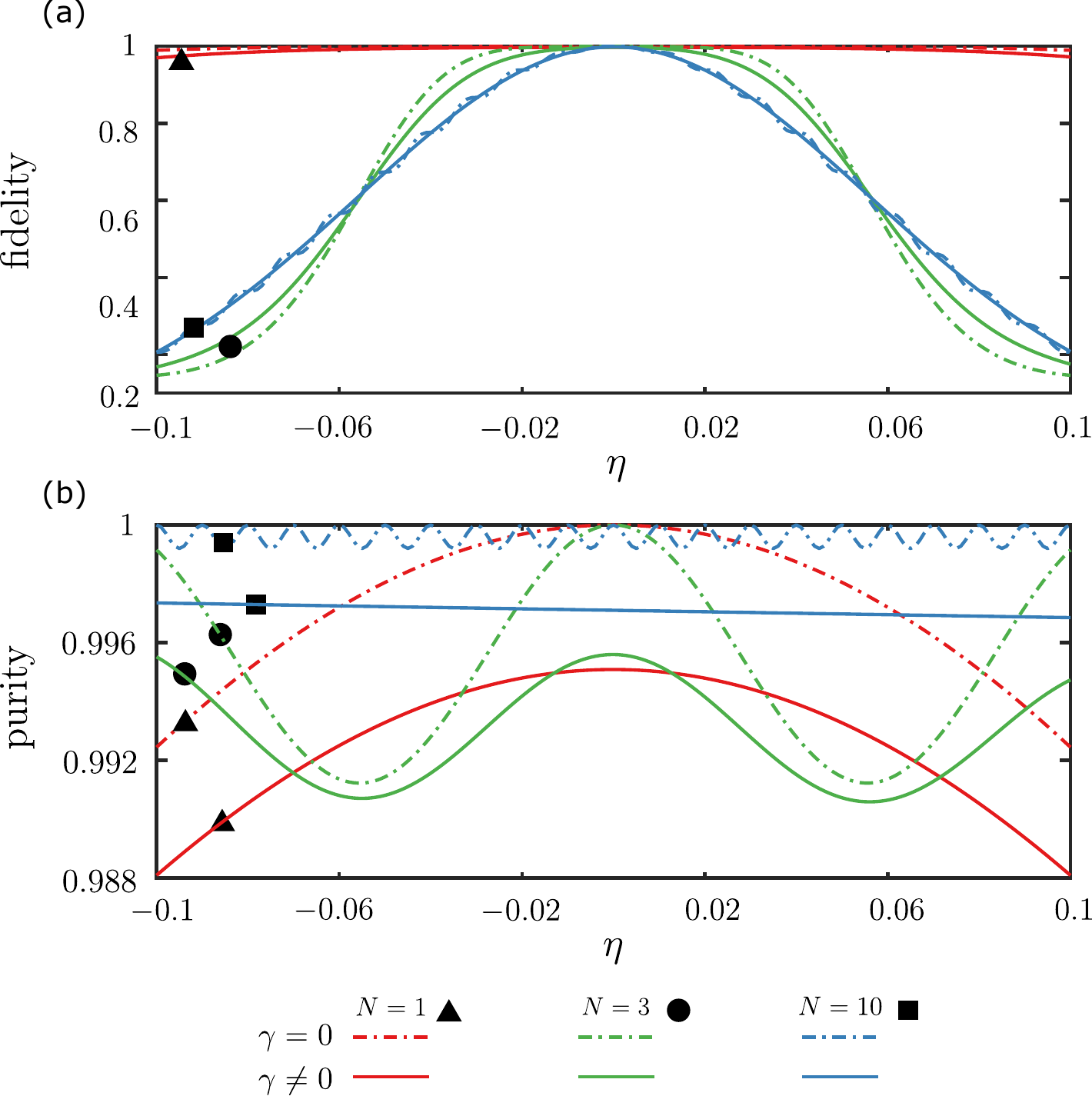}
\caption{Performance of S\o rensen-M\o lmer gate (continuous optomechanical interaction) including the dissipation of the mechanical oscillator, described by Eq.~\eqref{eq:master_OM}, for $\alpha=100$, $k=0.001$, $n_{\mathrm{th}}=0$, and $\gamma/\omega_m=0.02$. (a) Fidelity of the final field state as a function of $\eta$ for different $N$. (b) Purity of the final field state as a function of $\eta$ for different $N$. Legends are the same as in Fig.~\ref{fig:m_dsp}.}\label{fig:s_dsp}
\end{figure}

If the cavity photon decay rate is large $(\kappa>\omega_m)$, the optomechanical system is in the pulsed interaction regime~\cite{vanner2011pulsed,armata2016quantum,pikovski2012probing}. To be specific, the field enters the cavity and interacts with the mechanical oscillator only for a very short time. After that the field exits the cavity via a delay line such that both the field and the mechanical oscillator evolve freely without the optomechanical interaction~\cite{pikovski2012probing}. Then the field re-enters the cavity from the delay line to interact with the mechanical oscillator, which is a repetition of the initial step, see Fig.~\ref{fig:om}(b). This process in the frame rotating with the field frequency is described by the unitary evolution operator
\begin{align}\label{eq:pulseOMunitary}
&\hat{U}^{(\mathrm{OM})}_{p,1}(N_p,\Delta t)=\exp[i \lambda \hat{a}^{\dagger}\hat{a}(\hat{b}^{\dagger}+\hat{b})/\sqrt{2}]\exp[-i\omega_m \Delta t\hat{b}^{\dagger}\hat{b}]\nonumber\\
&\times\exp[i \lambda \hat{a}^{\dagger}\hat{a}(\hat{b}^{\dagger}+\hat{b})/\sqrt{2}]\exp[-i\omega_m \Delta t\hat{b}^{\dagger}\hat{b}]\cdots\nonumber\\
&\times\exp[i \lambda \hat{a}^{\dagger}\hat{a}(\hat{b}^{\dagger}+\hat{b})/\sqrt{2}].
\end{align}
Here $\lambda=2\pi g_0/\kappa$ is the effective coupling strength. The term $\exp[i \lambda \hat{a}^{\dagger}\hat{a}(\hat{b}^{\dagger}+\hat{b})/\sqrt{2}]$ is repeated $N_p$ times. It describes the pulsed interaction between the field and the mechanical oscillator, where the free evolution of the mechanical oscillator is neglected due to the short time duration. The term $\exp[-i\omega_m \Delta t\hat{b}^{\dagger}\hat{b}]$ is repeated $N_p-1$ times. It describes the free evolution of the mechanical oscillator in between two pulsed interactions with the field, with $\Delta t$ describing the time interval between the two pulsed interactions. Note that $\Delta t\gg 2\pi/\kappa$. It is straightforward to show that $\hat{U}^{(\mathrm{OM})}_{p,1}(N_p,\Delta t)=\hat{U}_p\times\exp[-i\omega_m(N_p-1)\Delta t\hat{b}^{\dagger}\hat{b}]$, where $\hat{U}_p$ is in the form of Milburn gate Eq.~\eqref{eq:pulse} with $\hat{O}=\hat{a}^{\dagger}\hat{a}$, and $\theta=2K_i\pi+\omega_m \Delta t$ with $K_i$ an integer. For simplicity, we will assume $K_i=0$. The term $\exp[-i\omega_m(N_p-1)\Delta t\hat{b}^{\dagger}\hat{b}]$ has no effect on the dynamics if the initial state of the mechanical oscillator is a thermal state.

Fig.~\ref{fig:4} shows the behaviour of the pulsed optomechanical interaction, taking the initial field state as a coherent state $|\alpha=100\rangle_f$, effective coupling strength $\lambda=0.001$, and thermal phonon number $n_{\mathrm{th}}=100$. The target gate is formed by $N_p=6$ and $\theta=2\pi/N_p$, corresponding to the mechanical phase space trajectory as a regular hexagon. As discussed in Sec.~\ref{sec:m}, suppose there is a relative error in the phase angle increment, $\theta'=(1+\xi)\theta$, which corresponds to a small error in controlling the time interval $\Delta t$ between two pulsed interactions. We consider the change in the gate performance when reducing $\lambda$ to $\lambda/N$, increasing $N_p$ to $N^2N_p$, keeping $\theta'$ unchanged. The results in Fig.~\ref{fig:4} clearly supports the arguments in Sec.~\ref{sec:m}. $\exp(i\hat{\psi}_{m,N})$ entangles the optical field with the mechanical oscillator, causing a blurred peak of the Q-function [Fig.~\ref{fig:4}(a2)-(a4)] and reducing the purity [Fig.~\ref{fig:4}(c)]. Its effect reduces as $N$ increases. $\exp(i\hat{\psi}_{O,N})$ represents an extra self-Kerr interaction, resulting in a rotation of the Q-function peak. The amount of rotation decreases as $N$ increases [Fig.~\ref{fig:4}(a)], leading to a larger fidelity [Fig.~\ref{fig:4}(b)].

It is worth pointing out that, for both S\o rensen-M\o lmer gate (Fig.~\ref{fig:3}) and Milburn gate (Fig.~\ref{fig:4}), when $N$ is finite, the fast oscillations of the fidelity and purity as a function of the relative error implies that, the corresponding curve does not stay above (or below) the curve of $N=1$ for all values of the error. However, in the limit $N\rightarrow\infty$, the fast oscillation is smoothened. The two curves only intersect at the point with zero error. We can thus unambiguously define whether the gate performance is improved by the transformation independent of the value of the error.

\subsection{Including the mechanical dissipation}\label{sec:dissipation}
Now we take into account an additional factor, which is the dissipation of the mechanical oscillator. This is relevant in the following way. An important property of both S\o rensen-M\o lmer gate and Milburn gate is that, under unitary dynamics, the state of the field is periodically disentangled with the mechanical oscillator. Therefore the properties of the mechanical oscillator will not affect the state of the field at these times. However, nonunitary dynamics of the mechanical oscillator will break the periodic disentanglement of the two modes, which are worth investigations.

The mechanical dissipation is included via the following master equation in Lindblad form~\cite{bose1997preparation},
\begin{equation}\label{eq:master_OM}
    \frac{d\rho(t)}{dt}=-\frac{i}{\hbar}[\hat{H}_u,\rho(t)]+\frac{\gamma}{2}\left(2\hat{b}\rho(t)\hat{b}^{\dagger}-\hat{b}^{\dagger}\hat{b}\rho(t)-\rho(t)\hat{b}^{\dagger}\hat{b}\right),
\end{equation}
where $\rho(t)$ is the joint state of the optical field and the mechanical oscillator, $\hat{H}_u$ is the Hamiltonian corresponding to the unitary dynamics, and $\gamma$ is the rate of the mechanical dissipation. In writing down the master equation, we have assumed that the mechanical oscillator is in contact with a vacuum bath. For simplicity, we only consider the case where the initial state is a product state of the optical field in a coherent state and the mechanical oscillator in a vacuum state, namely,
\begin{equation}
    \rho(0)=|\alpha\rangle_f\langle\alpha|\otimes|0\rangle_m\langle0|.
\end{equation}
Same as before, we assume $\alpha$ is real.

Let us start from Milburn gate, which is the pulsed interaction regime. The unitary dynamics without dissipation is described by Eq.~\eqref{eq:pulseOMunitary}. The mechanical dissipation only affects the steps where the mechanical oscillator evolves freely (see Fig.~\ref{fig:om}(b)), namely, the $\exp[-i\omega_m \Delta t\hat{b}^{\dagger}\hat{b}]$ terms in Eq.~\eqref{eq:pulseOMunitary}. This is because the interaction time in the step of the pulsed optomechanical interaction (Fig.~\ref{fig:om}(b) step 1), $2\pi/\kappa$, is much smaller than the mechanical oscillator free evolution time $\Delta t$ (Fig.~\ref{fig:om}(b) step 2). Making use of the transformation of coherent state basis under the mechanical dissipation terms in Eq.~\eqref{eq:master_OM}~\cite{walls1985analysis}, the state of the system after $N_p$ pulsed optomechanical interactions and $N_p-1$ intervals containing mechanical dissipation is calculated to be
\begin{subequations}\label{eq:OM_pul}
    \begin{align}
    &\rho_{N_p}=e^{-\alpha^2}\sum_{l_1,l_2=0}^{\infty}\frac{\alpha^{l_1+l_2}}{\sqrt{l_1!l_2!}}A_{N_p-1}(l_1,l_2)R_{N_p}(l_1,l_2)\nonumber\\
    &\times|l_1\rangle_f\langle l_2|\otimes|il_1\Phi_{N_p}\rangle_m\langle il_2\Phi_{N_p}|,\\
    &\Phi_{N_p}=\frac{\lambda}{\sqrt{2}}\frac{1-e^{(-i\omega_m\Delta t-\frac{\gamma\Delta t}{2})N_p}}{1-e^{-i\omega_m \Delta t-\frac{\gamma \Delta t}{2}}},\\
    &D=1-2e^{-\gamma \Delta t/2} \cos(\omega_m \Delta t)+e^{-\gamma \Delta t},\\
    &A_{N_p-1}(l_1,l_2)=\exp\Big[-\frac{\lambda^2}{4}(l_1-l_2)^2(1-e^{-\gamma \Delta t})\nonumber\\
    &\times\frac{1}{D}\big(N_p-1+\frac{e^{-\gamma \Delta t}(1-e^{-(N_p-1)\gamma\Delta t})}{1-e^{-\gamma\Delta t}}\nonumber\\
    &-\frac{2}{D}\big(e^{-\frac{\gamma \Delta t}{2}}\cos(\omega_m\Delta t)-e^{-\gamma \Delta t}-e^{-\frac{\gamma}{2}N_p\Delta t}\cos(N_p\omega_m \Delta t)\nonumber\\
    &+e^{-\frac{\gamma}{2}(N_p+1)\Delta t}\cos((N_p-1)\omega_m\Delta t)\big)\big)\Big],\\
    &R_{N_p}(l_1,l_2)=\exp\Big[i\frac{\lambda^2}{2}(l_1^2-l_2^2)\Big(\frac{(N_p-1)e^{-\frac{\gamma\Delta t}{2}}\sin(\omega_m\Delta t)}{D}\nonumber\\
    &-\frac{1}{D^2}\big(e^{-\gamma\Delta t}\sin(2\omega_m \Delta t)-e^{-\frac{\gamma}{2}(N_p+1)\Delta t}\sin((N_p+1)\omega_m\Delta t)\nonumber\\
    &-2e^{-\frac{3}{2}\gamma\Delta t}\sin(\omega_m\Delta t)+2e^{-\frac{\gamma}{2}(N_p+2)\Delta t}\sin(N_p\omega_m\Delta t)\nonumber\\
    &-e^{-\frac{\gamma}{2}(N_p+3)\Delta t}\sin((N_p-1)\omega_m\Delta t)\big)\Big)\Big],
    \end{align}
\end{subequations}
where the optical field state is expressed in Fock state basis, the mechanical oscillator state is expressed in coherent state basis.
Note that the exponent of $A_{N_p-1}(l_1,l_2)$ is purely real while the exponent of $R_{N_p}(l_1,l_2)$ is purely imaginary. The state of the optical field is given by taking the partial trace over the mechanical oscillator,
\begin{equation}
    \rho_{f,N_p}=\mathrm{Tr}_m(\rho_{N_p}).
\end{equation}
As considered in the previous subsection, we assume that there is a small error in controlling the interval time between two pulsed optomechanical interactions, namely, $\Delta t=(1+\xi)2\pi/N_p\omega_m$, with the relative error $|\xi|\ll1$. We look into the performance of the gate if we reduce the interaction strength by a factor of $N$, $\lambda\rightarrow\lambda/N$, increase the number of pulse by a factor of $N^2$, $N_p\rightarrow N^2N_p$, without changing $\Delta t$. We use the fidelity of the gate and the purity of the optical state to characterise the gate performance. The expressions are listed in Appendix~\ref{app:dissipation}. To visualise the results, we choose the initial coherent state amplitude of the optical field as $\alpha=100$, the effective optomechanical coupling strength $\lambda=0.001$, the number of pulses $N_p=6$, and the rescaled mechanical dissipation rate $\gamma/\omega_m=0.02$. The fidelity and purity as a function of the relative error $\xi$ for different values of the factor $N$ are plotted in Fig.~\ref{fig:m_dsp}. We can see that, for the chosen parameters, increasing the factor $N$ improves both the fidelity and the purity, even after the dissipation of the mechanical oscillator is considered. For the fidelity of the gate, the presence of mechanical dissipation slightly smooths the oscillation of the fidelity as a function of the error $\xi$. For the purity of the final optical state, the presence of mechanical dissipation reduces the purity for each value of $\xi$, and also smooths the oscillation of the purity as a function of $\xi$. Note that as we have taken the initial state of the mechanical oscillator as a vacuum state, the purity is much closer to $1$ compared with the case in the previous subsection. We have also chosen not to show the Q-function, as the difference between different values of $N$ turns out to be invisible.

S\o rensen-M\o lmer gate, or equivalently, the continuous optomechanical interaction regime, can be derived by taking the continuous limit of Milburn gate, making use of the idea of Trotterization that is implicitly applied in Ref.~\cite{bose1997preparation}. To be specific, we first set both the pulsed interaction time $2\pi/\kappa$ and the interval time $\Delta t$ to a small time step $dt$. Then we take the limit $dt\rightarrow0$, $N_p\rightarrow\infty$, keeping the continuous interaction time $t\equiv N_p\cdot dt$ finite. These two steps transform the system from the pulsed interaction regime to the continuous interaction regime. In other words, the continuous interaction regime is described by inserting the Hamiltonian in Eq.~\eqref{eq:OM_hamiltonian} into $\hat{H}_u$ in the master equation Eq.~\eqref{eq:master_OM}. The continuous time dynamics are decomposed into a series of infinitesimal time steps, each one with length $dt$. Within each $dt$, the optomechanical interaction, the mechanical oscillator free evolution and the mechanical dissipation happen simultaneously (see Fig.~\ref{fig:om}(a)). However, as $dt$ is infinitesimal, the three processes can be separated into two sequential steps. The first one only contains the optomechanical interaction. The second one contains both the mechanical oscillator free evolution and the mechanical dissipation. This separation leads to a similar structure as Milburn gate (see Fig.~\ref{fig:om}(b)). Note that, the difference with Milburn gate is that, for Milburn gate, the time for the mechancial oscillator free evolution and the mechanical dissipation, labelled as $\Delta t$, is finite.

For the continuous interaction regime, the state of the system at time $t$ is calculated by taking the continuous limit of Milburn gate,
\begin{subequations}\label{eq:OM_cont}
    \begin{align}
    &\rho(t)=e^{-\alpha^2}\sum_{l_1,l_2=0}^{\infty}\frac{\alpha^{l_1+l_2}}{\sqrt{l_1!l_2!}}A_t(l_1,l_2)R_t(l_1,l_2)\nonumber\\
    &\times|l_1\rangle_f\langle l_2|\otimes|il_1\Phi_t\rangle_m\langle il_2\Phi_t|,\\
    &\Phi_t=\frac{g_0 }{\sqrt{2}(i\omega_m+\frac{\gamma}{2})}(1-e^{-(i\omega_m+\frac{\gamma}{2})t}),\\
    &A_t(l_1,l_2)=\exp\Big[\frac{g_0^2}{4\omega_m^2+\gamma^2}(l_1-l_2)^2\big(e^{-\gamma t}-1-\gamma t\nonumber\\
    &+\frac{4\gamma^2 e^{-\frac{\gamma t}{2}}}{\gamma^2+4\omega_m^2}(e^{\frac{\gamma t}{2}}-\cos\omega_mt+\frac{2\omega_m}{\gamma}\sin\omega_mt)\big)\Big],\\
    &R_t(l_1,l_2)=\exp\Big[i\frac{2g_0^2}{4\omega_m^2+\gamma^2}(l_1^2-l_2^2)\big(\omega_mt\nonumber\\
    &-e^{-\frac{\gamma t}{2}}\frac{4\omega_m^2-\gamma^2}{4\omega_m^2+\gamma^2}\sin\omega_m t-\frac{4\omega_m\gamma}{4\omega_m^2+\gamma^2}(1-e^{-\frac{\gamma t}{2}}\cos\omega_m t)\big)\Big],
    \end{align}
\end{subequations}
where same as before, the optical field is expanded in Fock state basis, and the mechanical oscillator is expanded in coherent state basis. Similar as before, we consider that there is a small relative error in the evolution time, $t=(1+\eta)2\pi/\omega_m$ for $|\eta|\ll 1$, causing the gate to be imperfect. We analyse how the gate performance changes if we reduce the optomechanical interaction strength $g_0\rightarrow g_0/N$ together with increasing the interaction time to $t=N^2(1+\eta)2\pi/\omega_m$, for an integer $N$. The analytical expressions are listed in Appendix~\ref{app:dissipation}. We show an example of the results by choosing the parameters in the following way. The amplitude of the initial coherent state of the field is $\alpha=100$, the dimensionless interaction strength is $k=g_0/\sqrt{2}\omega_m=0.001$, and the rescaled mechanical dissipation rate is $\gamma/\omega_m=0.02$. In Fig.~\ref{fig:s_dsp}, we plot the fidelity of the gate and the purity of the final optical state, as a function of the relative error $\eta$, for several values of $N$. For the chosen parameters, including mechanical dissipation does not change the qualitative responses of the gate to different values of $N$. Specifically, increasing the value of $N$ only improves the purity of the final optical field (Fig.~\ref{fig:s_dsp}(b)), not the fidelity of the gate (Fig.~\ref{fig:s_dsp}(a)). For each value of $N$, the comparison between the situations with and without mechanical dissipation is similar to the case of Milburn gate. Including mechanical dissipation smooths the oscillations of the fidelity as a function of the relative error $\eta$. For the purity of the final optical state, mechanical dissipation has two effects. One is to reduce the purity for each value of $\eta$. The other is to smooth the oscillation of the purity as a function of $\eta$.

The unification of the pulsed interaction scheme and the continuous interaction scheme is clearly demonstrated in the example of an optomechanical system. On the one hand, we obtain the result for the continuous interaction case, Eq.~\eqref{eq:OM_cont}, by taking the continuous limit of the pulsed interaction [see Eq.~\eqref{eq:OM_pul}]. This does not involve solving differential equations, as opposed to the method in Ref.~\cite{bose1997preparation}. On the other hand, the different behaviors of the two gates in the presence of the relative error in time, as plotted in Fig.~\ref{fig:m_dsp} and Fig.~\ref{fig:s_dsp} where the mechanical dissipation is included, can be unified by including the relative error in the Milburn gate interaction strength Eq.~\eqref{eq:int_strength}. We further comment on this in Appendix~\ref{app:dissipation} where analytical expressions are provided.

\section{conclusion}

Both S\o rensen-M\o lmer gate and Milburn gate are geometric phase gates on a target mode via interaction with one auxiliary mechanical oscillator mode. We show that S\o rensen-M\o lmer gate is the continuous limit of Milburn gate, including a geometrical explanation in the mechanical phase space. Both gates have the property that, if the mechanical phase space trajectory is closed, the two modes disentangle, thus the mechanical mode is only virtually involved. However, performances of the gates are reduced in the presence of error in gate parameters. We explicitly consider error in time for S\o rensen-M\o lmer gate and in phase angle increment for Milburn gate. The transformation of decreasing the interaction strength together with increasing the number of loops traversed in the mechanical phase space can reduce the entanglement between the target mode and the mechanical oscillator mode, thus increasing the purity of the target mode. It increases the fidelity of Milburn gate, but the fidelity of S\o rensen-M\o lmer gate depends on the competition between thermal effect of the mechanical mode and error-induced additional self-interaction. We point out that the difference is because the interaction strength becomes dependent on the relative error when taking the continuous limit, and once this dependence is taken into account, the behaviours of the two gates are understood in a single platform. We quantitatively illustrate this unification via an optomechanical system, where in addition we include the effect of the mechanical oscillator dissipation to emphasize the application of our unified framework.

\acknowledgements  

This work is supported by the KIST Open Research Program, the QuantERA ERA-NET within the EUs Horizon 2020 Programme, the UK Hub in Quantum Computing and Simulation with funding from UKRI EPSRC grant EP/T001062/1, and the EPSRC (EP/R044082/1) and the Royal Society. YM is supported by the EPSRC Centre for Doctoral Training on Controlled Quantum Dynamics at Imperial College London (EP/L016524/1) and funded by the Imperial College President's PhD Scholarship.


%

\newpage
\onecolumngrid
\appendix

\section{Analytical expressions of S\o rensen-M\o lmer gate and Milburn gate not included in the main text}~\label{app:0}

The explicit expressions of $c_1$, $c_2$ and $c_3$ in Eq.~\eqref{eq:pulse} are
\begin{subequations}
\label{eq:m}
\begin{align}
&c_1=\lambda\{\frac{1}{2}+\frac{1}{2}\cos[(N_p-1)\theta]+\frac{1}{2}\sin[(N_p-1)\theta]\cot(\frac{\theta}{2})\},\\
&c_2=\lambda\{\frac{1}{2}\cot(\frac{\theta}{2})[1-\cos((N_p-1)\theta)]+\frac{1}{2}\sin[(N_p-1)\theta]\},\\
&c_3=\frac{1}{2}\lambda^2\frac{N_p\sin\theta-\sin(N_p\theta)}{4\sin^2(\theta/2)},
\end{align}
\end{subequations}

The expression of S\o rensen-M\o lmer gate with relative error $\eta$ in interaction time, for $N=1$ (see Sec.~\ref{sec:s}), is
\begin{subequations}
\begin{align}
&\hat{U}_{c,N=1}(\eta)=\hat{V}_{m,N=1}\hat{V}_{O,N=1}\hat{U}_{c,T},\\
&\hat{V}_{m,N=1}=\exp\{i\sqrt{2}k\hat{O}[\sin(\eta2\pi)\hat{x}_m-(1-\cos(\eta2\pi))\hat{p}_m]\},\\
&\hat{V}_{O,N=1}=\exp\{ik^2\hat{O}^2[\eta2\pi-\sin(\eta2\pi)]\},\\
&\hat{U}_{c,T}=\exp(ik^2\hat{O}^22\pi).
\end{align}
\end{subequations}

\begin{figure}[h]
\centering
\includegraphics[width=0.5\textwidth]{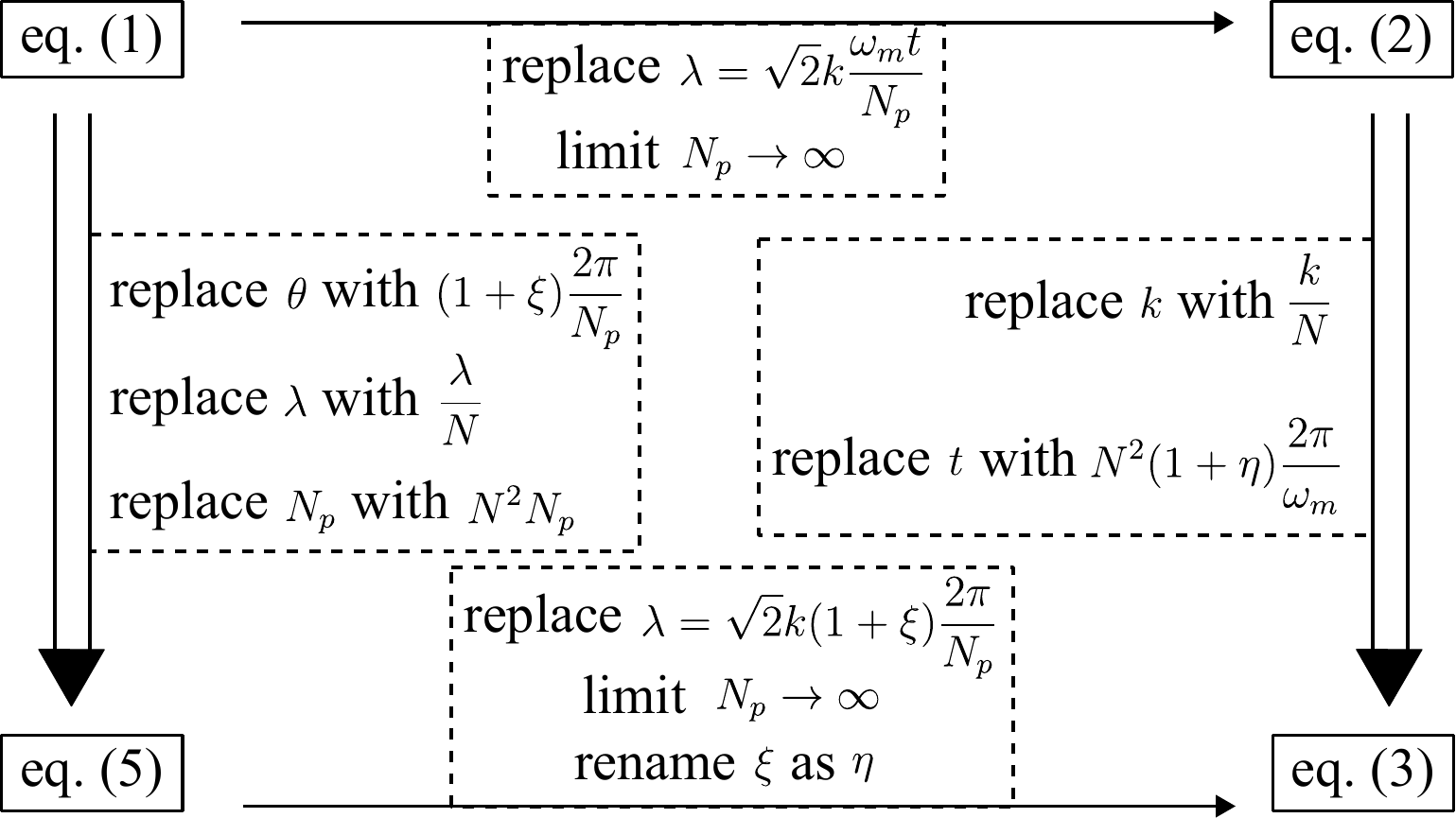}
\caption{The relations between the expressions in the pulsed interaction regime and the continuous interaction regime.}\label{fig:eq_relation}
\end{figure}

In Fig.~\ref{fig:eq_relation}, we show how the expressions in the main text are connected to each other. Double-arrow refers to an equality, while single-arrow refers to a limit. This figure represents the unified mathematical framework of the two gates.

\section{Analytical expressions of Q-function, fidelity and purity for the unitary evolution of the optomechanical example}\label{app:unitary}

For the continuous interaction regime, the Q-function of the field state at time $t'=N^2(1+\eta)2\pi/\omega_m$ is (for real $\alpha$)

\begin{subequations}
\begin{align}
&Q_c(\beta)=\frac{1}{\pi}\sum_{l_1,l_2=0}^{\infty}e^{-\alpha^2-|\beta|^2}\frac{\alpha^{l_1+l_2}\beta^{*l_1}\beta^{l_2}}{l_1!l_2!}\Psi_c(l_1,l_2)M_c(l_1,l_2),\\
&\Psi_c(l_1,l_2)=\exp\{i\frac{k^2}{N^2}(l_1^2-l_2^2)[(1+\eta)2\pi N^2-\sin(\eta2\pi N^2)]\},\\
&M_c(l_1,l_2)=\exp\{-\frac{k^2}{N^2}(l_1-l_2)^2(2n_{\mathrm{th}}+1)[1-\cos(\eta 2\pi N^2)]\}.
\end{align}
\end{subequations}
The fidelity is
\begin{equation}
F_c=\sum_{l_1,l_2=0}^{\infty}e^{-2\alpha^2}\frac{\alpha^{2(l_1+l_2)}}{l_1!l_2!} \exp[-i2\pi k^2(l_1^2-l_2^2)]\Psi_c(l_1,l_2)M_c(l_1,l_2).
\end{equation}
The purity is
\begin{equation}
P_c=\sum_{l_1,l_2=0}^{\infty}e^{-2\alpha^2}\frac{\alpha^{2(l_1+l_2)}}{l_1!l_2!}M_c^2(l_1,l_2).
\end{equation}

For the pulsed interaction regime, the Q-function of the field for phase angle increment $\theta'=(1+\xi)2\pi/N_p$ is
\begin{subequations}
\begin{align}
&Q_p(\beta)=\frac{1}{\pi}\sum_{l_1,l_2=0}^{\infty}e^{-\alpha^2-|\beta|^2}\frac{\alpha^{l_1+l_2}\beta^{*l_1}\beta^{l_2}}{l_1!l_2!}\Psi_p(l_1,l_2)M_p(l_1,l_2),\\
&\Psi_p(l_1,l_2)=\exp\{i\frac{\lambda^2}{N^2}(l_1^2-l_2^2)\frac{N^2N_p\sin[(1+\xi)2\pi/N_p]-\sin(N^22\pi\xi)}{8\sin^2[(1+\xi)\pi/N_p]}\},\\
&M_p(l_1,l_2)=\exp\{-\frac{\lambda^2}{N^2}(l_1-l_2)^2(2n_{\mathrm{th}}+1)\frac{1-\cos(\xi2\pi N^2)}{8\sin^2[(1+\xi)\pi/N_p]}\}.
\end{align}
\end{subequations}
The fidelity is
\begin{equation}
F_p=\sum_{l_1,l_2=0}^{\infty}e^{-2\alpha^2}\frac{\alpha^{2(l_1+l_2)}}{l_1!l_2!}\exp[-i\lambda^2(l_1^2-l_2^2)\frac{N_p}{4}\cot(\frac{\pi}{N_p})]\Psi_p(l_1,l_2)M_p(l_1,l_2).
\end{equation}
The purity is
\begin{equation}
P_p=\sum_{l_1,l_2=0}^{\infty}e^{-2\alpha^2}\frac{\alpha^{2(l_1+l_2)}}{l_1!l_2!}M_p^2(l_1,l_2).
\end{equation}

\section{Analytical expressions of fidelity and purity for the optomechanical example including mechanical dissipation}\label{app:dissipation}

For the continuous interaction regime, the fidelity for interaction time $t=N^2(1+\eta)2\pi/\omega_m$ is
\begin{subequations}
\begin{align}
    &\tilde{F}_c=\sum_{l_1,l_2=0}^{\infty}e^{-2\alpha^2}\frac{\alpha^{2(l_1+l_2)}}{l_1!l_2!}\exp[-i2\pi k^2(l_1^2-l_2^2)]\mathcal{P}_c(l_1,l_2)\mathcal{M}_c(l_1,l_2),\\
    &\mathcal{P}_c(l_1,l_2)=\exp\Big[i\frac{4k^2\omega_m^2}{4\omega_m^2+\gamma^2}(l_1^2-l_2^2)\big((1+\eta) 2\pi-\frac{1}{N^2}e^{-\frac{\gamma}{\omega_m}\pi(1+\eta)N^2}\frac{4\omega_m^2-\gamma^2}{4\omega_m^2+\gamma^2}\sin(\eta 2 \pi N^2)\nonumber\\
    &\ \ \ -\frac{1}{N^2}\frac{4\omega_m\gamma}{4\omega_m^2+\gamma^2}(1-e^{-\frac{\gamma}{\omega_m}\pi(1+\eta)N^2}\cos(\eta 2\pi N^2))\big)\Big],\\
    &\mathcal{M}_c(l_1,l_2)=\exp\Big[-\frac{k^2}{N^2}(l_1-l_2)^2\frac{4\omega_m^2}{4\omega_m^2+\gamma^2}\big(1-\cos(\eta 2\pi N^2)e^{-\frac{\gamma}{\omega_m}\pi(1+\eta)N^2}+\frac{\gamma}{\omega_m}\pi(1+\eta)N^2\nonumber\\
    &\ \ \ -\frac{2\gamma^2}{\gamma^2+4\omega_m^2}e^{-\frac{\gamma}{\omega_m}\pi(1+\eta)N^2}(e^{\frac{\gamma}{\omega_m}\pi(1+\eta)N^2}-\cos(\eta 2\pi N^2)+\frac{2\omega_m}{\gamma}\sin(\eta 2 \pi N^2))\big)\Big].
    \end{align}
\end{subequations}
The purity is
\begin{align}
     &\tilde{P}_c=\sum_{l_1,l_2=0}^{\infty}e^{-2\alpha^2}\frac{\alpha^{2(l_1+l_2)}}{l_1!l_2!}\mathcal{M}_c^2(l_1,l_2).
\end{align}
Note that here we have chosen $n_{\mathrm{th}}=0$.

For the pulsed interaction regime, the fidelity for phase angle increment $\theta=(1+\xi)2\pi/N_p$ is
\begin{subequations}
\begin{align}
    &\tilde{F}_p=\sum_{l_1,l_2=0}^{\infty}e^{-2\alpha^2}\frac{\alpha^{2l_1+2l_2}}{l_1!l_2!}\exp[-i\lambda^2(l_1^2-l_2^2)\frac{N_p}{4}\cot(\frac{\pi}{N_p})]\mathcal{P}_p(l_1,l_2)\mathcal{M}_p(l_1,l_2),\\
    &\mathcal{D}=1-2e^{-\gamma (1+\xi)2\pi/2N_p\omega_m} \cos((1+\xi)\frac{2\pi}{N_p})+e^{-\gamma (1+\xi)2\pi/N_p\omega_m},\\
    &\mathcal{P}_p(l_1,l_2)=\exp\Big[i\frac{\lambda^2}{2N^2}(l_1^2-l_2^2)\times\Big(\frac{1}{\mathcal{D}}(N^2N_p-1)e^{-\gamma(1+\xi)2\pi/2N_p\omega_m}\sin((1+\xi)\frac{2\pi}{N_p})-\frac{1}{\mathcal{D}^2}\times\big(e^{-\gamma(1+\xi)2\pi/N_p\omega_m}\sin((1+\xi)\frac{4\pi}{N_p})\nonumber\\
    &\ \ \ -e^{-\frac{\gamma}{2}(N^2N_p+1)(1+\xi)2\pi/N_p\omega_m}\sin(N^22\pi\xi+(1+\xi)\frac{2\pi}{N_p})-2e^{-\frac{3}{2}\gamma(1+\xi)2\pi/N_p\omega_m}\sin((1+\xi)\frac{2\pi}{N_p})\nonumber\\
    &\ \ \ +2e^{-\frac{\gamma}{2}(N^2N_p+2)(1+\xi)2\pi/N_p\omega_m}\sin(N^22\pi\xi)-e^{-\frac{\gamma}{2}(N^2N_p+3)(1+\xi)2\pi/N_p\omega_m}\sin(N^22\pi\xi-(1+\xi)\frac{2\pi}{N_p})\big)\Big)\Big],\\
    &\mathcal{M}_p(l_1,l_2)\nonumber\\
    &=\exp\Big[-\frac{\lambda^2}{4N^2}(l_1-l_2)^2\big(\frac{1}{\mathcal{D}}(1-e^{-\gamma (1+\xi)2\pi/N_p\omega_m})\big(N^2N_p-1\nonumber+\frac{e^{-\gamma (1+\xi)2\pi/N_p\omega_m}(1-e^{-(N^2N_p-1)\gamma(1+\xi)2\pi/N_p\omega_m})}{1-e^{-\gamma(1+\xi)2\pi/N_p\omega_m}}\nonumber\\
    &-\frac{2}{\mathcal{D}}\big(e^{-\gamma (1+\xi)2\pi/2N_p\omega_m}\cos((1+\xi)\frac{2\pi}{N_p})-e^{-\gamma (1+\xi)2\pi/N_p\omega_m}-e^{-\frac{\gamma}{2}N^2N_p(1+\xi)2\pi/N_p\omega_m}\cos(N^22\pi\xi)+e^{-\frac{\gamma}{2}(N^2N_p+1)(1+\xi)2\pi/N_p\omega_m}\nonumber\\
    &\times\cos(N^22\pi\xi-(1+\xi)\frac{2\pi}{N_p})\big)\big)+\frac{1}{\mathcal{D}}(1-2e^{-\gamma N^2 N_p (1+\xi)2\pi/2N_p\omega_m}\cos(N^22\pi\xi)+e^{-\gamma N^2 N_p (1+\xi) 2\pi/N_p\omega_m})\big)\Big]
\end{align}
\end{subequations}
The purity is
\begin{align}
    &\tilde{P}_p=\sum_{l_1,l_2=0}^{\infty}e^{-2\alpha^2}\frac{\alpha^{2l_1+2l_2}}{l_1!l_2!}\mathcal{M}_p^2(l_1,l_2).
\end{align}

The unification of the two gates in the presence of the mechanical dissipation can be shown in the following way. We insert Eq.~\eqref{eq:int_strength} into $\mathcal{P}_p(l_1,l_2)$ and take the limit $N_p\rightarrow\infty$ to arrive at $\mathcal{P}_c(l_1,l_2)$. Similarly, we insert Eq.~\eqref{eq:int_strength} into $\mathcal{M}_p(l_1,l_2)$ and take the limit $N_p\rightarrow\infty$ to arrive at $\mathcal{M}_c(l_1,l_2)$. The conversion from the purity $\tilde{P}_p$ to $\tilde{P}_c$ is therefore straightforward. For the fidelity $\tilde{F}_p$, in addition to the transforms of $\mathcal{P}_p(l_1,l_2)$ and $\mathcal{M}_p(l_1,l_2)$, we replace $\lambda$ in the exponential factor $\exp[-i\lambda^2(l_1^2-l_2^2)\frac{N_p}{4}\cot(\frac{\pi}{N_p})]$ with $\sqrt{2}k\cdot2\pi/N_p$. Note that this does not include the relative error $\xi$, as the exponential factor comes from the target state. Taking the limit $N_p\rightarrow\infty$ brings $\tilde{F}_p$ to $\tilde{F}_c$.

\end{document}